\documentclass[twocolumn,showpacs]{revtex4}  
\usepackage{amsmath, amsthm, amssymb}
\usepackage{graphicx,wrapfig,lipsum}
\usepackage{dcolumn}  
\usepackage{bm}       
\usepackage{float}
\usepackage{subcaption}
\usepackage{hyperref}
\usepackage{enumerate}
\usepackage{xcolor}
\usepackage{verbatim}

\begin{document}
\widetext
\leftline{\textit{Entropy} \textbf{2020}, 22, 312; doi:10.3390/e22030312}
\title{A Maximum Entropy Method for the Prediction of Size Distributions}
\author{Cornelia Metzig$^{1,2}$, Caroline Colijn$^{1,3}$\\
\textit{$^{1}$Imperial College London, UK, $^{2}$Queen Mary University London, UK, $^{3}$Simon Fraser University, Vancouver, Canada}\\
\tt{c.metzig@imperial.ac.uk, ccolijn@sfu.ca}
}
\begin{abstract}
We propose a method to derive the stationary size distributions of a system, and the degree distributions of networks, using maximisation of the Gibbs-Shannon entropy. We apply this to a preferential attachment-type algorithm for systems of constant size, which contains exit of balls and urns (or nodes and edges for the network case). Knowing mean size (degree) and turnover rate, the power law exponent and exponential cutoff can be derived. Our results are confirmed by simulations and by computation of exact probabilities.
We also apply this entropy method to reproduce existing results like the Maxwell-Boltzmann distribution for the velocity of gas particles, the Barabasi-Albert model and multiplicative noise systems.
\end{abstract}

%\keyword{complex networks; growth process; fluctuation scaling; Gibbs-Shannon entropy; scalefree~distribution}

\pacs{02.50.Fz, 05.10.Gg, 05.40.-a, 05.65.+b}
\maketitle
\section{Introduction}
The famous model by Yule \cite{yule1925mathematical,simon1955class} and its analogue for networks, the Barabasi and Albert (BA) model for scalefree networks \cite{barabasi1999emergence}, have been widely used to the describe phenomena and processes that involve scalefree distributions. 
The latter are an ubiquitous phenomenon found, for example, in word frequency in language \cite{mandelbrot1953informational} and web databases \cite{babbar2014power}, city and company sizes \cite{axtell2001zipf} and high-energy physics, and they have been modeled with different approaches, for example, References \cite{marsili1998interacting,biro2005power}.
When occurring in the degree distribution of networks, power laws affect in particular the dynamics on a network, for example, of protein interaction networks \cite{albert2005scale}, brain functional networks \cite{eguiluz2005scale}, email networks \cite{ebel2002scale}, and various social networks \cite{barabasi2000scale} such as respiratory contact networks \cite{eubank2004modelling}. 
An advantage of the the Yule model and the BA-model is that their interpretation of the 'preferential attachment' process (in which nodes preferentially attach to existing nodes with high degree) is simple and plausible, and that they generate a scalefree degree distribution, whose exponent can be calculated analytically given the rate of introduction of nodes. Therefore simple preferential attachment continues to be widely used to simulate networks for spreading processes. In addition,  it has been extended \cite{bertotti2019bass, bertotti2019configuration, bhaumik2018conserved, glos2019spectral, jaiswal2018evocut} and the process has been generalized \cite{courtney2018dense}. The exponent of the degree distribution in the BA-model can be derived starting from a master equation \cite{newman2005power}. This ansatz is solvable for constantly growing systems, but becomes too complicated when a system can also lose nodes and edges. However, continuous growth is often not fulfilled in real world examples, especially for social systems, because people also exit the system or network. 

Here, we present a method to predict the scaling exponent and the exponential cutoff of a size/degree distribution by maximisation of the Gibbs-Shannon entropy, building on the work in Reference \cite{jaynes1957information}. This method is applicable to a variety of models that do not require the hypothesis of continuous growth. We introduce it at the example of a micro-founded model for the size distribution of urns (filled with balls), which preserves a stationary size distribution by deletion of balls, and/or by deletion of urns. Like the Yule process, this algorithm can be extended to networks, where links and nodes are entering and exiting the network. 

Our example model also explains another scaling phenomenon, a `tent-shaped' probability density for the aggregate growth rate $g_t$, which often occurs in combination with a scalefree distribution in many real-world examples \cite{halvarsson2019asymmetric, alves2013scaling, schwarzkopf2010explanation, bottazzi2006explaining, takayasu2014generalised, bottazzi2002corporate, alfarano2012statistical, thornhill2003learning, picoli2006scaling, stanley1996scaling, amaral1997scaling, coad2007firm}. Tent-shaped growth rate probabilities are also generated by other preferential-attachment models like BA, but they are not produced by other families of models for scalefree distributions. 

\section{A Preferential-Attachment Algorithm for A Stable Size Process}
We consider a system of urns and balls, and extend it to nodes and edges in Section \ref{sec:networks}. Each~of the $M$ urns is filled with $n_i$ balls, and their sizes sum to a variable number $\sum_{i=1}^M n_i=N$. The~dynamics are framed in terms of urns receiving and losing balls, in discrete time steps $k$. The key features are that $M$ is constant; the average of $N$ is conserved over time; the expected value of size change, $\left\langle n_{i,t+1}-n_{i,t}\right\rangle$ for individual urns is $0$ (though $n_i$ is not stationary), and that every ball has the same chance of attracting another ball and of vanishing. 
We give now the succession of events in one iteration (where we refer to $N_t$ as the number of balls \textit{at the beginning} of an iteration). A scheme with more details is given in Figure \ref{fig:scheme_iteration} in Appendix \ref{sec.sup.scheme}.
\begin{itemize}
 \item[1.] Growth of urns: every ball has probability $q_t$ of attracting another ball from a reservoir. 
 Let $X_{i,t}$ be the number of new balls in urn $i$ in iteration $t$; $X_{i,t}$ is binomial with mean $n_{i,t}q_t$, such that the urn grows on average to $n_{i,t}(1+q_t)$. To ensure that after a full iteration of steps 1--4 $N_t$ is stationary (but not fixed), after the first time step, $q_t$ is adjusted to $q_t=\frac{N_{t=0}}{N_t}(1+q_{t=0})-1$, such that the expected size \textit{after} growth is always $N_{t=0}(1+q_{t=0})$.
 \item[2.] Shrinking of urns: every ball has probability $\delta_{shrink, t}$ of disappearing, which is adjusted to  $\delta_{shrink, t}=\sum_iX_{i,t}/(N_t+\sum_iX_{i,t})$ as a result of the growth step 1, such that the expectancy after shrinking $\left\langle N \right \rangle$ equals the initial size $N_{t=0}$. Let $Y_{i,t}$ be the number of disappearances of urn $i$; $Y_{i,t}$ is a random variable with a binomial distribution with mean $\left\langle Y_{i,t} \right \rangle =\delta_{shrink}(n_{i,t}+X_{i,t})$.
The system shrinks in the number of balls, and some urns might be be left with $0$ balls (which can be interpreted as exiting urns).
 \item[3.] Exit of urns (and balls): every urn has probability $\delta_{exit}$ of exiting, that is, being set to size $0$, so the system loses balls.
 \item[4.] Entry of urns (and balls): Urns that have lost all their balls due to steps (2) or (3) are replaced by urns that contain $1$ ball, so that $M$ is strictly conserved after one iteration of steps 1--4, but $N_t$ fluctuates around $N_{t=0}$.
\end{itemize}

Even if step 3 is omitted, some urns will exit, as urns can vanish by losing all their balls. Steps 3 and 4 do not affect the number of urns $M$, but may leave the system with a net loss or gain of balls, compared to the beginning of step 1. To conserve the average sum of balls in the system \textit{after} growth, $\left\langle N_t+\sum_iX_{i,t}\right \rangle$, the probability $q_t$ to attract a new ball from the reservoir is adjusted for the next iteration. A scheme of how the number of urns and balls change in steps 1--4 is given in Figure \ref{fig:scheme_iteration} in Appendix~\ref{sec.sup.scheme}.
 \subsection{Possible Cases}
 \label{sec:possible_cases}
 This general process can be reduced to two limiting scenarios with the same growth but different shrinking mechanisms. These are: (I) No deletion of urns of size $n >0$. 
 (II) Urns can only grow and do not shrink, but exit (with their balls) at a rate $\delta_{exit}$ and get replaced by urns of size $1$. 
 (III) A combination of both.
 \begin{itemize}
 \item [(I)]
 Urns do not exit (step 3 is omitted), that is, $\delta_{exit}=0$.
For an urn $i$ of size $n_i$, the probability distribution of the size after a growth-and-shrink cycle, $p(n_{i,t+1}|n_{i,t})$ can be written as a discrete Gaussian centered around $n_{i}$ and with standard deviation
 \begin{equation}
 \sigma(n_i)=\left(\frac{q}{(1+q)^2} \, 2\,n_i \right)^{\omega}\equiv (\hat q \,2n_i)^{\omega},
 \label{eq:sigma_scaling_gauss}
  \end{equation}
with standard deviation scaling exponent $\omega=0.5$ (see Equations (\ref{eq:proba_nt2})--(\ref{eq:proba_nt0}) in Appendix \ref{sec:sup.discrete}). 

\item [(II)] Urns do not shrink (step 2 is omitted). At each step a fraction $\delta_{exit}$ of urns is deleted and replaced by urns of size $1$, which means that the number of exiting balls varies more strongly. 
With~probability $\delta_{exit}$, the urn size in the next time step is $1$ (if one thinks of the replacing urn as being the same urn); with probability $1-\delta_{exit}$, the urn grows by $X_i$, and the binomial distribution of $X_i$ has standard~deviation
\begin{equation}
\sigma(n_{i})=\left(q(1-q)n_{i}\right)^{\omega},
\label{eq:sigma_scaling_binomial}
\end{equation}
again with scaling exponent $\omega=0.5$. The most probable outcome is the maximum of the binomial distribution (unless this is lower than $\delta_{exit}$), but that is \textit{not} the average expected size at $t+1$ for an urn of size $n_i$:  $\left\langle n_{i,t+1} \right\rangle$ is again precisely $n_{i,t}$ (see Figure \ref{fig:entropy_with_turnover} 
{left} column with examples).
\item [(III)] Mixed case. 
 Steps 2 and 3 can be combined such that some balls (a fraction $\delta_{shrink}$) will disappear from the system due to shrinking of urns, and some because urns exit with probability $\delta_{exit}$ with their balls. 
 The exiting urns have on average the median size of all urns in the system, on average a fraction $\delta_{exit, balls,t}$ of balls exits with them (with $\delta_{exit, balls,t} = \delta_{exit, urns}\cdot \frac{\text{median urn size}}{\text{mean urn size}}$). The turnover rate can then be defined as the fraction of balls that gets removed through exit of urns, normalized by the total number of balls that get removed in one time step,
 $\mu=\frac{\delta_{exit, balls,t}}{\delta_{exit, balls,t}+\delta_{shrink}}$. 
 \end{itemize}

\section{Maximum Entropy Method}
\label{sec:entropy}

To derive $P_n$ we use the Maximum Entropy Principle (MEP) \cite{jaynes1957information,rosenkrantz1989we}, which can be defined as:
\begin{quote}
Suppose that probabilities are
to be assigned to the set of $k$ mutually exclusive possible outcomes of an
experiment. The principle says that these probability values are to be chosen
such that the (Shannon) entropy of the distribution $P = (P_1, \ldots , P_k)$, that is, the expression $S(P)=-\sum_n P_n \log P_n$ attains its maximum value under the condition that $P$ agrees with the given
information  \cite{uffink1996constraint}.\end{quote} We use as given information the probabilities $p(n_{i,t+1}|n_{i,t})$ of every urn $i$ to change size. 
For urns that do not exit, the probability $p(n_{i,t+1}|n_{i,t})$ is either Gaussian (case I) or binomially distributed (case II), and their associated entropies are approximated by $s=\frac{1}{2} \ln (2 \pi \sigma^2)$. This term becomes $s_i=\frac{1}{2} \ln (2 \pi \, 2\hat q\, n_i)$ for case (I) using (\ref{eq:sigma_scaling_gauss}), or $s_i=\frac{1}{2} \ln (2 \pi \, 2 q(1-q)\, n_i)$  for case (II) using (\ref{eq:sigma_scaling_binomial}), which both have a standard deviation that scales as $\sigma(n_i)\propto \sqrt{n_i}$. The individual Gaussian or binomial probability distributions $p(n_{i,t+1}|n_{i,t})$ are themselves maximum entropy distributions \cite{harremoes2001binomial}, under the constraint that the expectation of change $\left\langle n_{i,t+1}-n_{i,t}\right \rangle $ is zero. 
If the $s_i$ are maximal at stationary state, then $\sum_{i=1}^{M} s_i$ needs to be stationary. $p(n_{i,t+1}|n_{i,t})$ are generated by the same updating rules as $P_n$, and according to the Maximum Entropy Principle, $P_n$ needs to agree with the given information about $p(n_{i,t+1}|n_{i,t})$. (Should $\sum s_i$ not be relevant then it would cancel out in the entropy maximisation step, as it is the case for a system shown in Section \ref{sec:other_systems}). Formulated differently, the size distribution $P_n$ maximizes entropy under the constraint $\frac{1}{M}\sum_{i=1}^{M} s_i = C^*$. Subtracting the constant $\frac{1}{2} \ln (2 \pi \, 2 \hat q)$ from $C^*$, we can use as sum of entropies $s_i$
\begin{equation}
C= \frac{1}{M}\sum_{i=1}^{M} \ln(n_i)\,.
\label{eq:C_uncorrected}
\end{equation}

A second constraint is the conservation of the expectation of individual urn size $\left\langle n_{i,t+1} \right\rangle =\sum_{n_{i,t+1}} n_{i,t+1}p(n_{i,t+1}|n_{i,t}) =n_{i,t}$, or summed over all urns $i$, $\sum_i \sum_{n_{i,t+1}} n_{i,t+1}p(n_{i,t+1}|n_{i,t})  =\sum_i n_{i,t}$ which can be written as
$\sum_n P_{n} n =E$. (The precise number $N_t$ fluctuates around $N_0$, in analogy to the total energy of a gas described in the canonical ensemble.) 
The Lagrangian function for maximizing entropy of the urn size distribution is
\begin{equation*}
S(P) = \sum_n P_n \ln P_n + \lambda \left( \sum_n P_n \ln (n) - C\right)
\end{equation*}
\vspace{-0.5cm}
\begin{equation}
 +\beta\left(\sum_n P_n n -E\right).
 \label{eq:entropy}
\end{equation}
To determine the distribution that maximizes $S$, we calculate $\frac{\partial S}{\partial P_n}$ and set it to $0$, leading to 
\begin{equation}
P_n=K n^{-(\alpha+1)}e^{-\beta n},
\label{eq:powerlaw_analytical}
\end{equation}
with $\alpha+1=\lambda/2$.
This equation can be solved using $\sum_n P_n=1$, $\sum_n P_n \ln n =C$ and  $\sum_n P_n n =E$, which gives $C=\frac{K}{\beta^{2-\alpha}}\int_{a_0}^{\infty}dn \frac{\Gamma (2-\alpha, \beta n)}{n}$ and $E=\beta^{2-\alpha}\frac{\Gamma(2-\alpha,\beta)}{\Gamma(2-\alpha, \beta)}$ (with $\Gamma$ the upper incomplete Gamma function). 
For $\beta=0$ the constant in Equation (\ref{eq:powerlaw_analytical}) becomes $K=(\lambda-1)a_0^{\lambda-1}$, if urn sizes $n$ can take values in $[a_0, \infty)$. Knowing $K$, the exponent $\alpha +1$ can be determined from the condition $ \sum_n P_n \ln n = C$. In continuous approximation $ \int_{a_0}^{\infty} d n P_n \ln n = C$ this yields 
$\lambda=1+ \alpha=1+\frac{1}{C-\ln a_0}$.
This result is independent of $q$ and for $a_0=1$ simplifies to
\begin{equation}
\alpha =\frac{1}{C}\,.
\label{eq:exponent}
\end{equation}

For $\beta=0$, $\alpha$ depends only on $C$, which is the logarithm of the geometric mean of urn sizes.  
\section{Results}\vspace{-6pt}

\subsection{Size Distribution}
\begin{figure}%[H]
\centering
\begin{subfigure}[]{0.9\columnwidth}
\includegraphics[angle=0, width=0.85\columnwidth]{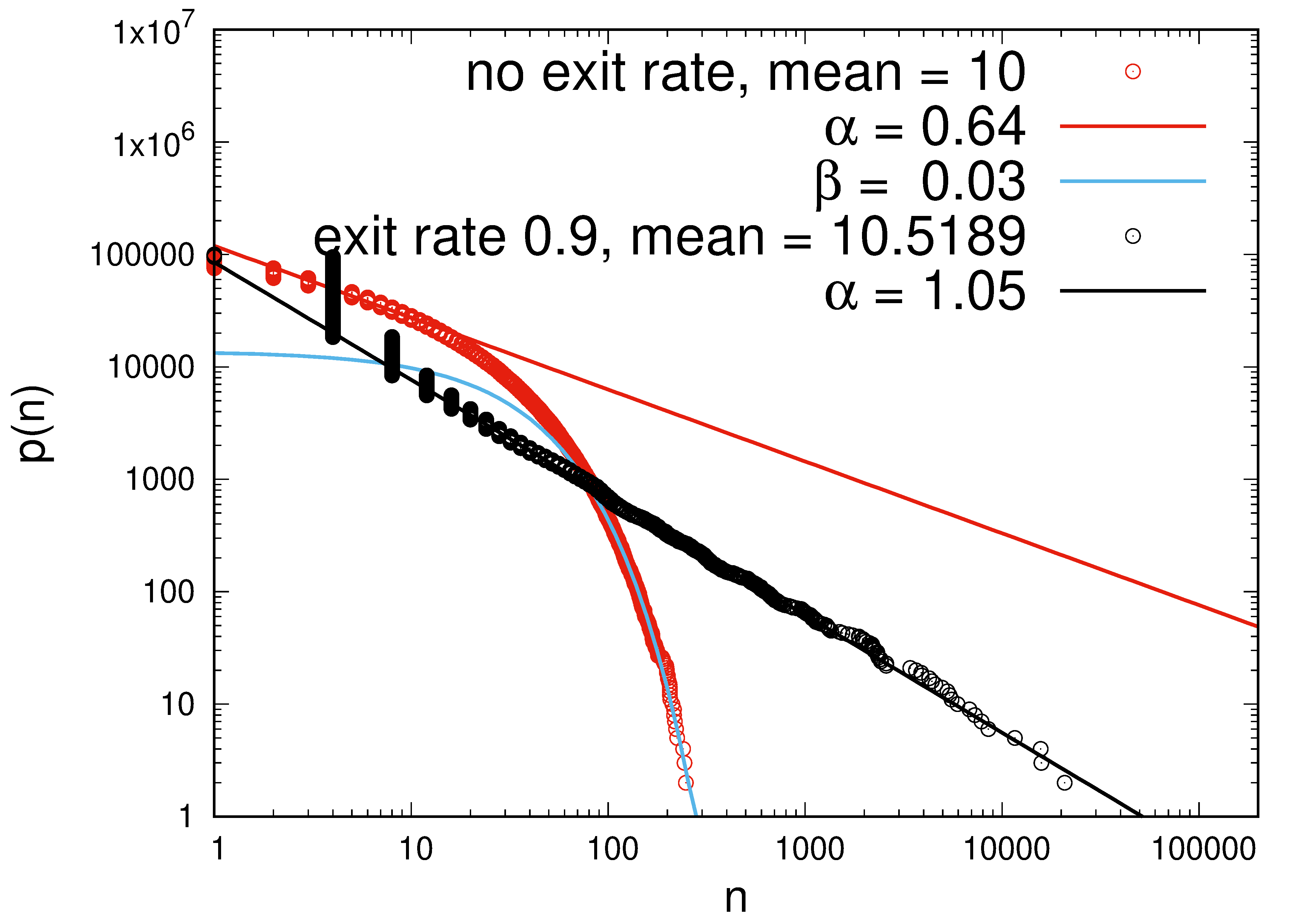}
\caption{}
\end{subfigure}
\hspace{0.5cm}
\begin{subfigure}[]{0.9\columnwidth}
\includegraphics[angle=0, width=0.8\columnwidth]{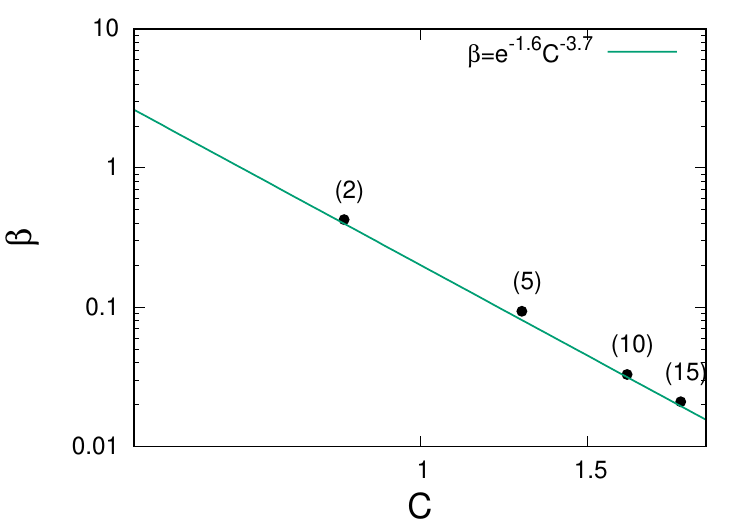}
\caption{}
\end{subfigure}
%\vspace{1cm}
\begin{subfigure}[]{0.9\columnwidth}
\includegraphics[angle=0, width=\columnwidth]{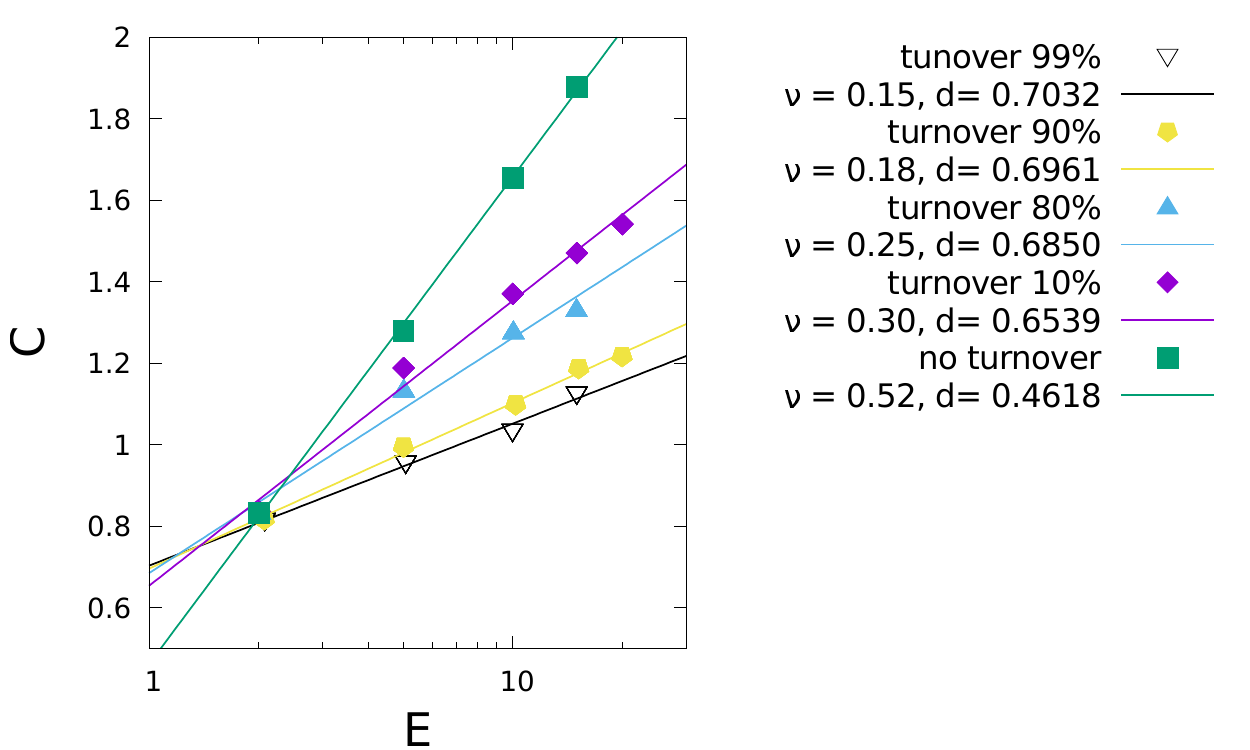}
\caption{}
\end{subfigure}
\hspace{0.5cm}
\begin{subfigure}[]{0.9\columnwidth}
\includegraphics[angle=0, width=\columnwidth]{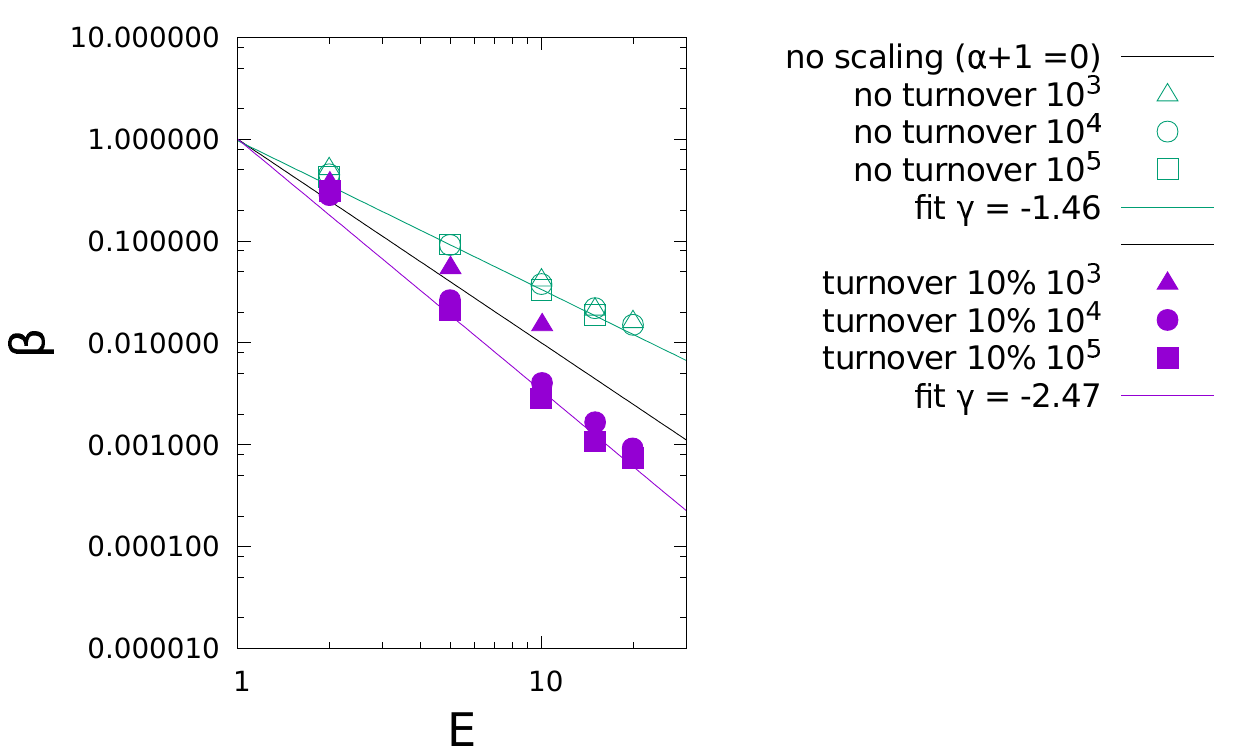}
\caption{}

\end{subfigure}

 	\caption{Simulation results for different turnover rates. For small system sizes, for high turnover rates $\mu$ and $E$ the cutoff is no longer a clear exponential which is why in subfigure (\textbf{d}) for $N=10^3$ some $\beta$ are lacking. (\textbf{a}) Example of size distributions for different exit rates, in double logarithmic scale. $\left\langle n\right\rangle=10$. (\textbf{b}) $\beta$ vs. $C$ for no turnover rate and different mean sizes in parentheses. (\textbf{c}) Numerical $C$ vs. mean $E$.  (\textbf{d}) Exponential cutoff $\beta$ vs. E.}
	\label{fig:Simulation_E_C_beta}
\end{figure}

\begin{figure}%[H]\centering
\includegraphics[angle=0, width=0.85\columnwidth]{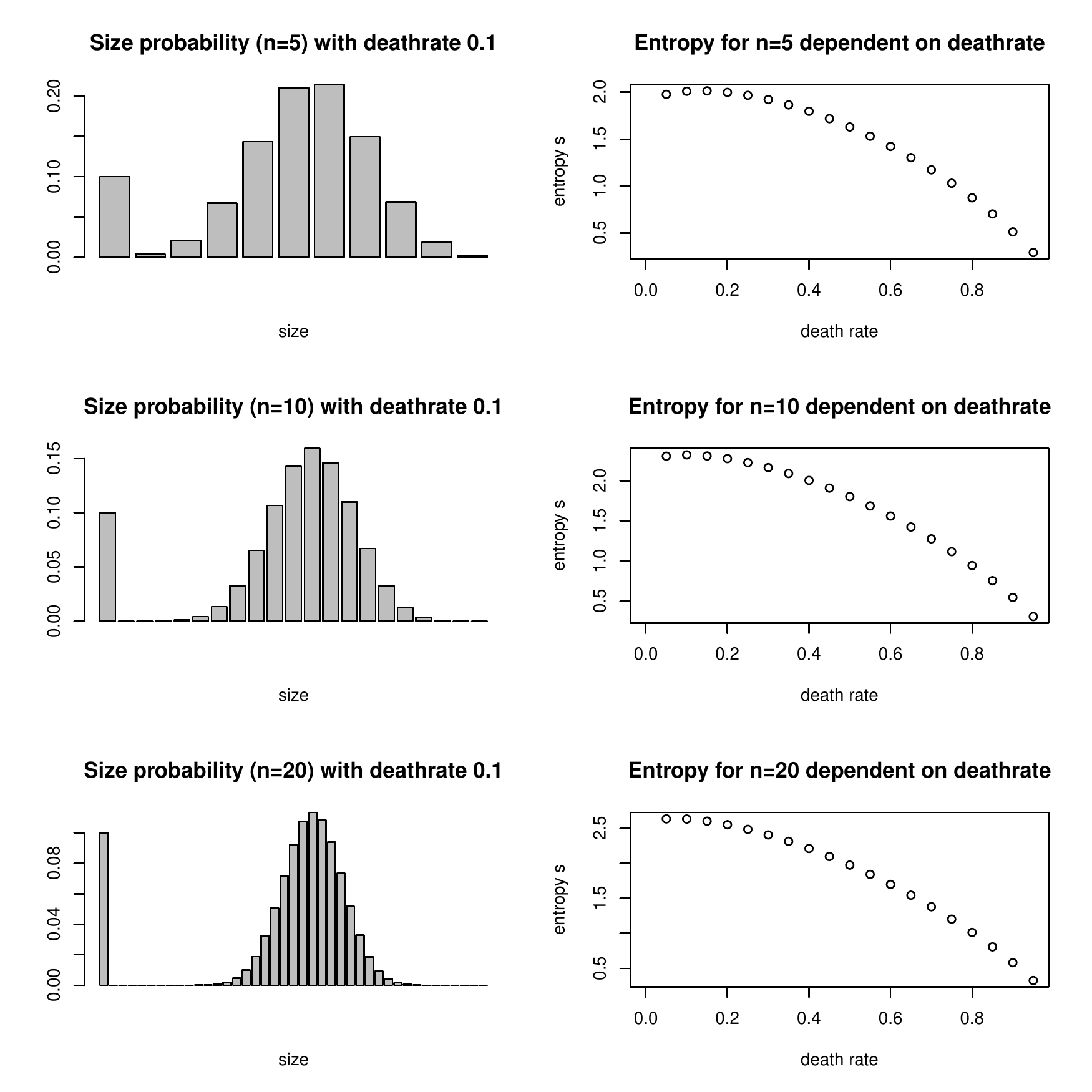}
\caption{Three examples for the expected size distribution, and its entropy, for the mixed case (III) with deathrate 0.1, for urns of size 5, 10 and 20. The high probability for size n=1 comes from the re-introduction of urns of size 1. Dependent on the deathrate, this probability increases and the sum of the other outcomes decreases. In the right column, the entropy $s_i=-\sum_i p_i\log pi $ is calculated as a function of the deathrate $\delta_{exit}$. The higher the deathrate, the lower $s$, which contributes to a lower $C=\sum_i s_i$, and a higher the exponent $\alpha$, which are shown in Figure \ref{fig:Simulation_E_C_beta}.
}
\label{fig:entropy_with_turnover}
\end{figure}

The maximum entropy size distribution of the stable size process (\ref{eq:powerlaw_analytical}) is confirmed by numerical results (see Figure \ref{fig:Simulation_E_C_beta}a). The method holds for all three cases. The sum $C$ is smaller when urns have a probability $\delta_{exit}$ to be replaced with an urn of size $1$, which has a theoretical explanation. $C=\frac{1}{M}\sum_{i=1}^{M} \ln n_i$ is the logarithm of the geometric mean of the urn sizes $\left\langle m \right\rangle_{geom}=\exp[\frac{1}{M}\sum_{i=1}^{M}\ln n_i]$. Reference \cite{mitchell_2004} has shown that the geometric mean decreases when subject to mean-preserving spread, that is, when all numbers in a set become proportionately farther from their fixed arithmetic mean (of course increasing their standard deviation). The statement applies to $C$ as well, since the logarithm is a monotonically increasing function. Spread increases with turnover, which increases the fraction of urns of size $1$, and consequently the larger the other urns need to be for a given arithmetic mean $E$. 
Another effect is the direct change of  $s_i$ through increasing $p(1|n_{i,t})$ with higher turnover, as shown in Figure \ref{fig:entropy_with_turnover}.

To compare the exponent fit of $\alpha$ to the calculated one from the numerical entropy sum $C$, an adjustment to the computation of $C$ in (\ref{eq:C_uncorrected}) is necessary, since the approximation of the entropy of a binomial $s(n)=\frac{1}{2} \ln (2 \pi q(1-q) n)+ \mathcal{O}(\frac{1}{n})$ holds for large $n$, but yields $s_{n=1}=0$. Especially for cases (II) and (III) with turnover, urns of size $1$ make up a large fraction of urns, and their contribution to the total entropy cannot be neglected. To correct for this we calculate the exact entropies $s_{e,n=1}$ and $s_{e,n=2}$ from the definition $s_e=\sum_i p_i \ln p_i$, and then multiply their fraction by $s_{n=2}$ from the large-n-approximation: $s_{n=1}=\frac{s_{e,n=1}}{s_{e,n=2}}s_{n=2}$ with $\frac{s_{e,n=1}}{s_{e,n=2}}=\frac{q\ln q+(1-q)\ln (1-q))}{q^2 \ln q^2+ 2 q(1-q\ln[2q(1-q)]+(1-q^2)\ln (1-q^2))}\approx 0.6 $ for a wide range of $q$. We use as corrected $C$
\begin{equation}
 C_{corr}=\frac{1}{N}\sum_n \ln n + \sum_{i, n_i=1} \frac{s_{e,n=1}}{s_{e,n=2}} s_{n=2}.
  \label{eq:corrected_C}
\end{equation}

\begin{figure}
\includegraphics[angle=0, width=0.6\columnwidth]{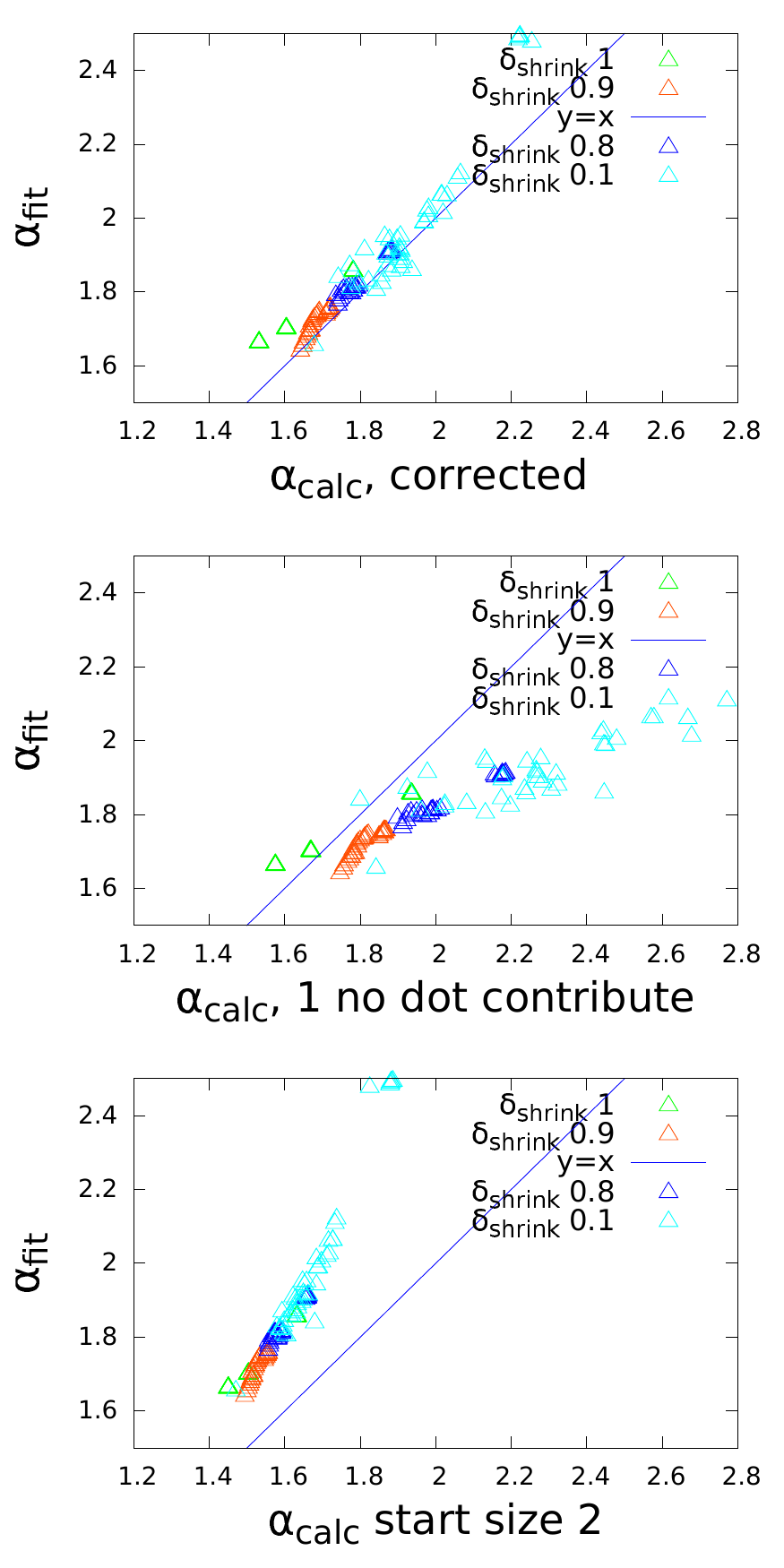}
 	\caption{Fitted vs. calculated exponent $\alpha$, for three different ways of accounting for urns of size $1$.}
	\label{fig:alpha_alpha_large}
\end{figure}

The correction is only significant for high turnover rates where a large fraction of urns has size $1$, and with it, the theoretical $\alpha$ is confirmed by simulations (see Figure \ref{fig:alpha_alpha_large}). Furthermore, if the average size $E$ and turnover rate $\mu$ are known, the power law exponent $\alpha$ (via the constant $C$) and the exponential decay $\beta$ can be determined numerically (see Figure \ref{fig:Simulation_E_C_beta}c,d). 
In case (I) where urns shrink, $C$ is so big and consequently $\alpha$ so low, that $P_n$ has a strong exponential cutoff in order to keep the system at the
same mean urn size, in agreement with (\ref{eq:powerlaw_analytical}). Although (\ref{eq:exponent}) holds only for $\beta=0$, it only slightly overestimates $\alpha$ for $\beta > 0$, since the exponential cutoff affects only a small fraction of urns. In the presence of $\beta>0$, $C$ can be greater than $1$, resulting in $\alpha <1$, which would diverge without exponential cutoff. 

The larger $\mu$ and the mean urn size $E$, the larger the fluctuations in number of removed balls in step $3$, and the more the urn size distribution fluctuates. 
Both $\alpha$ and $\beta$ are independent of system size (except if the system size is too low for convergence, in which case $\beta$ increases), see Figure \ref{fig:Simulation_E_C_beta}d. Simulation results are independent of the urns' probability to attract balls in one time step, $q$, in agreement with our theoretical result in (\ref{eq:exponent}).

In addition to simulations, we derived the same size distributions for cases (I) and (II) with another method using the exact probabilities of $p(n_t|n_{t-1})$ for every individual urn, which we calculated with a recursion equation (see Appendix \ref{sec:sup.discrete} and Figure \ref{fig:superposition_simulation}a). Also this method reproduces all of the results of the maximum entropy method which we presented above and in Section \ref{sec:other_systems} (see for example Figure~\ref{fig:superposition_simulation}b).

\subsection{Aggregate Growth Rate Distribution of the Stable Size Process}
\begin{figure}%[H]
	\centering
\includegraphics[angle=0, width=0.9\columnwidth]{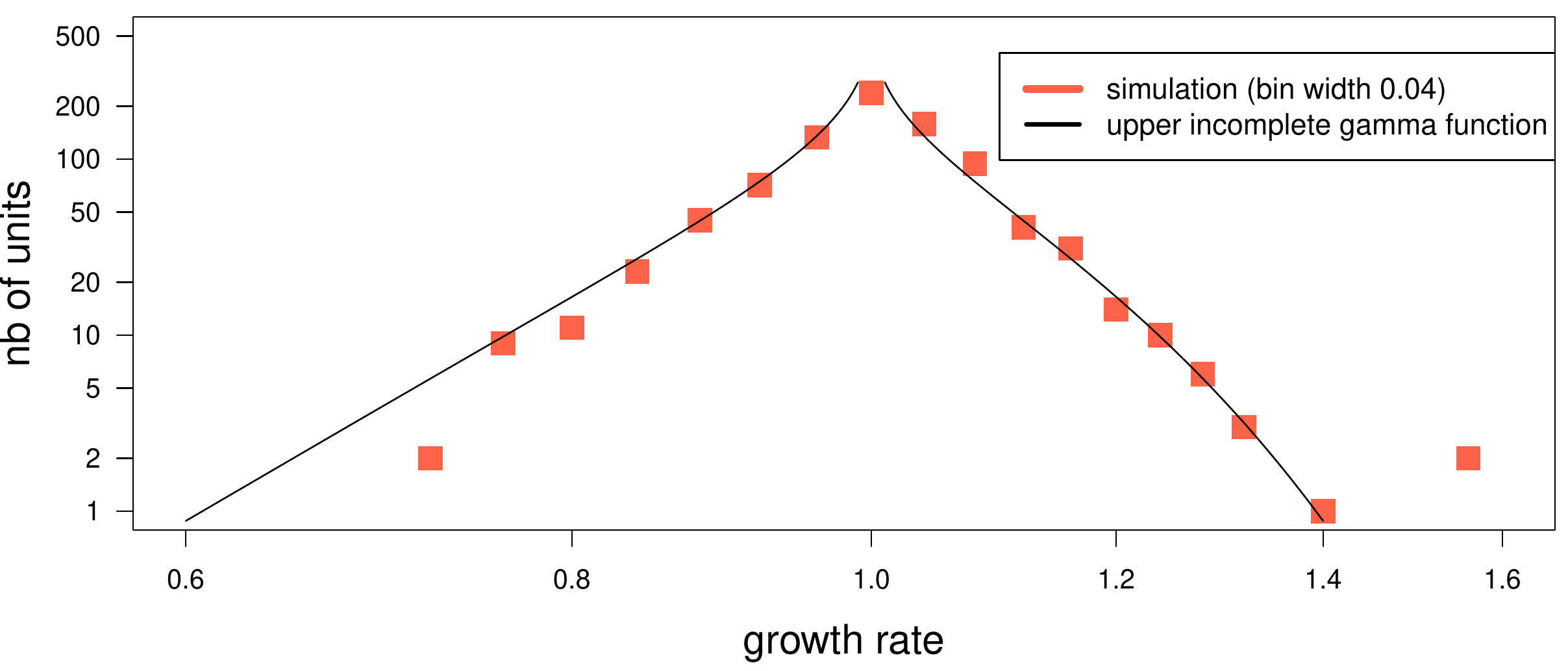}
\vspace{-0.15cm}
 	\caption{\small Aggregate growth rate distribution, simulation and fit (for $\beta=0$, $\alpha=0.5$)}
	\label{fig:upper_incomplete_gamma}
\end{figure}

It follows from the binomially (or normally) distributed $p(n_{i,t}|n_{i-t,1})$ (where $\sigma (n) \propto n^{0.5}$) that an \textit{individual} urn's growth rate, defined as $g_{i,t}=\frac{n_{i,t}}{n_{i,t-1}}$, is also normally distributed 
\begin{equation}
 \mathcal{G}(g_{i,t}|n_{i,t-1})=\sqrt{\frac{n_{i,t-1}}{2\pi\,c}} e^{-\frac{1}{2} \frac{n_i}{c}\, (g_{i,t}-1)^{2}},
\label{eq:gaussian_growthrate_independent}
\end{equation}
 with scaling $\sigma_g(n)\propto n^{-0.5}$. The \textit{aggregate} growth rate distribution (aggregated over all urns in one timestep, dropping the index $t$) is $ \mathcal{G}(g)=\sum_{i=1}^{N}p(n_i)\mathcal{G}(g_{i} | n_{i})$, or in the continuous limit $ {\mathcal{G}}(g)=\int_{n_0}^{\infty}d n\, \mathcal{G}(g|n)\rho(n)$.
This can be evaluated using (\ref{eq:gaussian_growthrate_independent}) and for $\rho(n)$ the expression (\ref{eq:powerlaw_analytical}). For $\alpha=0.5$ and $\beta=0$, this yields a upper incomplete Gamma function shown in Figure \ref{fig:upper_incomplete_gamma} and  References \cite{metzig2012heterogeneous,metzig2013model, metzig2014model}:  $\mathcal{G}(g)\propto \Gamma\small{\left(0, \frac{1}{2}n_{0}(g-1)^2\right)}$. 
Such `tent-shaped' aggregate growth rate distributions are often observed for quantities that themselves follow a power-law  \cite{picoli2006scaling, bottazzi2006explaining, stanley1996scaling, mondani2014fat, alfarano2012statistical}. The tent shape is the \textit{sample average}, but not the expectation for a given urn, as other models for it presume \cite{fu2005growth,schwarzkopf2010explanation}. This result adds credibility to the stable size process as a model for some real system, in particular since a tent-shaped aggregate growth rate distribution does not automatically result from other models for scalefree distributions. An example is a multiplicative noise term $\gamma$ in the linear Langevin equation $n_{t+1}= \gamma n_{t}+ \delta $  \cite{takayasu1997stable, biro2005power} (where $\delta$ is additive noise and $n_t$ is the size of the process at time $t$). Such models produce a scalefree distribution for $n$ above some value $n'$, but the growth rate $\gamma$ can be any i.i.d. random variable \cite{dorogovtsev2000scaling, sarshar2004scale, moore2006exact} independent of an urn's size $n$, and no distinction between individual growth rate and aggregate growth rate can be made. Therefore it does not additionally generate a tent shape for the aggregate growth rate distribution (unless a tent shape is assumed as individual growth rate distribution $\gamma$). 
\subsection{Extension to Networks of the Stable Size Process}\label{sec:networks}
The algorithm can be adapted to derive the degree distribution for networks, where $M$ nodes are connected with $N$ undirected and unweighted links. The substeps become:
(1. and 2.) A random link is broken, and one of its neighbors $i$ is chosen to receive an additional link (i.e., every node is picked with probability proportional to its degree $n_i$). Its new neighbor $j$ is also picked with probability $\propto n_j$. 
(3.) Nodes are removed at random at rate $\delta_{exit}$; their links are broken.
(4.) Nodes are re-introduced and linked to an existing node; the probability of selecting a node $i$ as neighbor is $\propto n_i$. New links are added to keep $N$ conserved; each node has a probability of receiving a link $\propto n_i$.

Compared to an urn/ball system, the exponential cutoff always exists, for the following reason. The case (II) in Section \ref{sec:possible_cases}, where the only shrink mechanism is exit nodes, cannot be reached. If a node exits the network, all its links are broken, so necessarily also non-exiting nodes will lose the same number of edges. The maximal turnover $\mu$ rate is therefore 0.5. Numerical results confirm that a scalefree network without cutoff is not produced by this algorithm.

\begin{figure}
\centering
\begin{subfigure}[]{0.3\linewidth}
\includegraphics[width=\columnwidth]{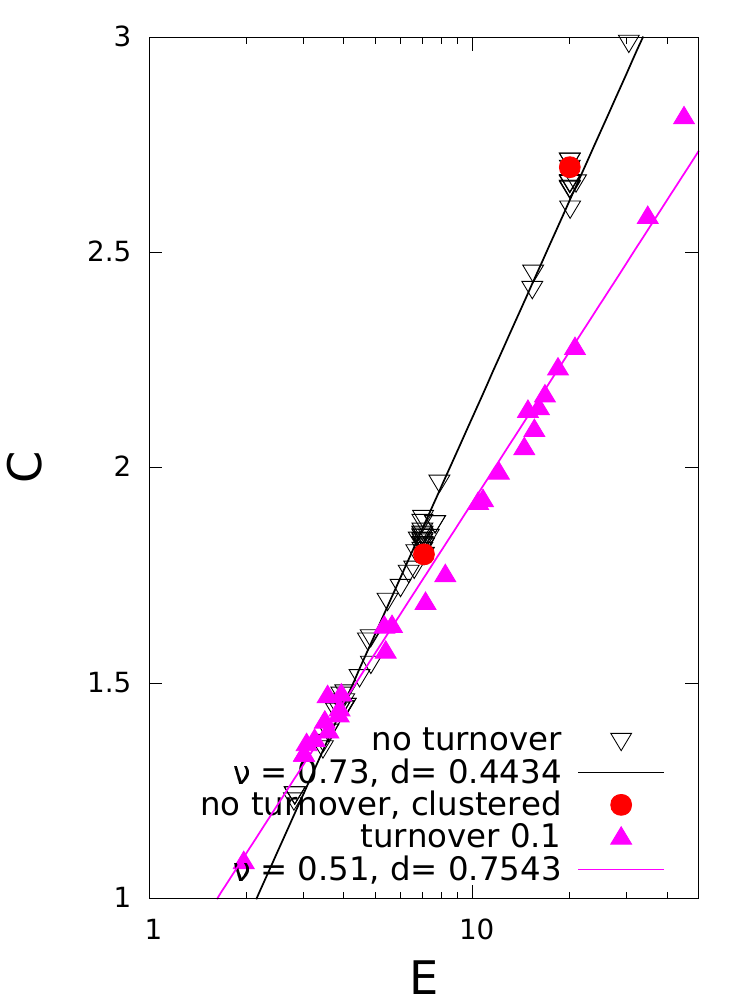}
\caption{}
\end{subfigure}
\begin{subfigure}[]{0.65\linewidth}
\includegraphics[width=\columnwidth]{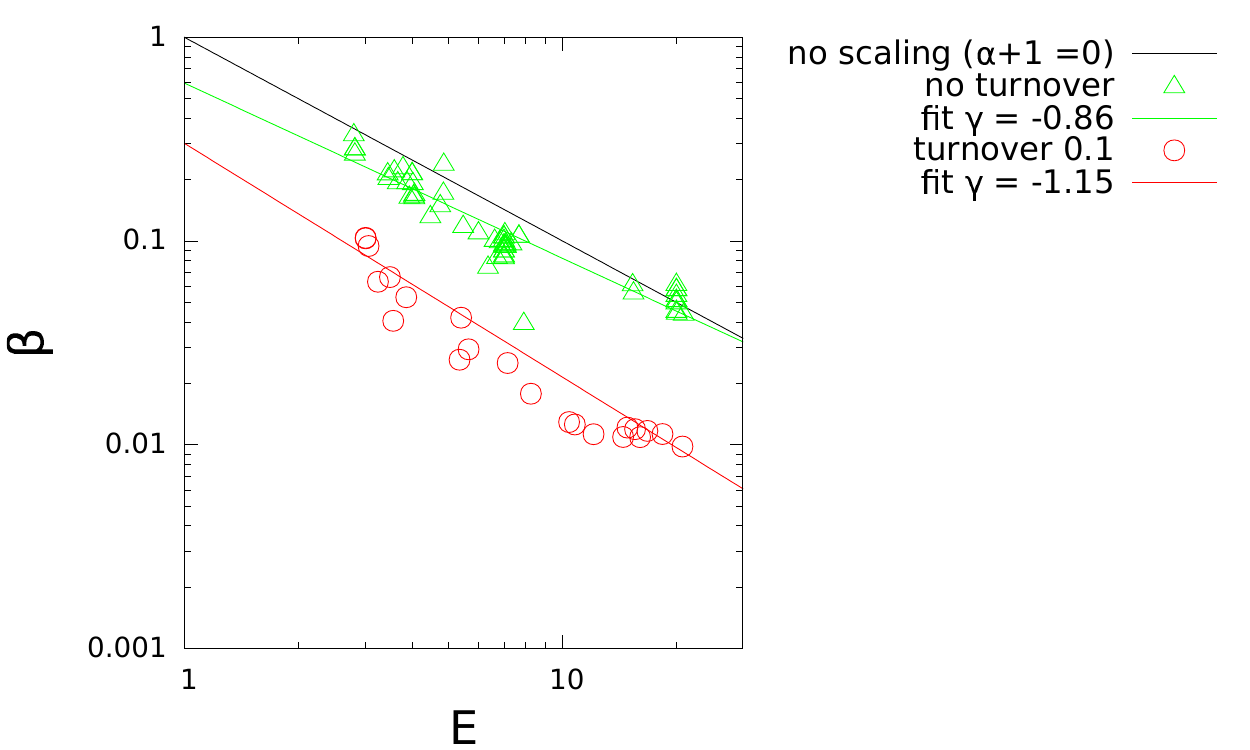}
\caption{}
\end{subfigure}
 	\caption{Simulated network and simulation results for different turnover rates, $N=10^3$. (\textbf{a}) Numerical $C$ vs. mean degree $E$.  (\textbf{b}) Exponential cutoff $\beta$ vs. E.}
	\label{fig:networks_three}
\end{figure}

In previous work \cite{metzig2017impact, metzig2019phylogenies}, we have added further features to make the model more plausible for example, for epidemiology, such as clustering (that a link is preferably formed between neighbours of second or third degree), or different exit rules, for example, removal of a node after a given time span instead of exit by rate $\delta_{exit}$. The latter increases in addition the exponential cutoff, because it prevents nodes to remain a sufficiently long duration to attract many links.
In that case $\alpha$ and $\beta$ in (\ref{eq:powerlaw_analytical}) can still be inferred numerically from $E$, $\mu$ and additional features (see Figure \ref{fig:networks_three}).

\subsection{Maximum Entropy Argument of Other Systems}
\label{sec:other_systems}
The method of using the sum of entropies of the evolution of individual urns as a constraint on the entropy of the system can be applied to many urn-ball systems in discrete time steps. It will affect the maximum entropy size distribution whenever $s_i$ depend on $n_i$.
\subsubsection{Maxwell-Boltzmann Distribution} A well-known example for a maximum entropy distribution is the velocity distribution of gas particles (Maxwell-Boltzmann distribution, here in one dimension). The only assumption about the process generating the velocity distribution $P$ is that the mean 
$E = \sum_i \epsilon_i \propto \sum_i v_i^2$ is conserved in time. Particles can change their velocity through collisions with other particles. In a given timespan, the sum of received shocks of particle $i$ (in one dimension) follows a Gaussian distribution, which has entropy $s_i=\frac{1}{2} \ln (2 \pi \sigma^2)$, but all particles are hit by shocks of the same distribution, that is, $s_i=s$, since $\sigma$ does not depend on a particle's own velocity $v_i$. The focus is usually not on the distribution of individual change of $v_i$, only on the stationary distribution of $v$. In one dimension, the Lagrangian function becomes $S(P) = \sum_v P_v \ln P_v + \lambda \left( \sum_v P_v s_v - C\right) +\beta\left(\sum_v P_v v^2 -E\right)$ with $\lambda=0$ at the extremum where $\frac{\partial S}{\partial v}=0$, and results in $P_v=K \exp(-\beta v^2$). The constraint on entropies vanishes, as $s_i$ do not depend on $v_i$, and the established Maxwell-Boltzmann distribution is found.
\subsubsection{Yule Process (or Barabasi-Albert for Networks)} We simulated the Yule process in discrete time steps of adding a number $N_{add}$ of balls before adding an urn. If we consider larger time steps where several urns and many balls are added, the growth of an urn is approximately binomial with $\omega=0.5$. 
Following our argument in Section \ref{sec:entropy}, the binomial $p(n_{i,t+1}|n_{i,t})$ are themselves maximum entropy distributions \cite{harremoes2001binomial} and therefore the sum of their entropies is stationary.
The system has the constraint that the sum of individual entropies $C=\frac{1}{M}\sum_{i=1}^M s_i$ in one time step is stationary. 
The Lagrangian function becomes
$S(P) = \sum_n P_n \ln P_n + \lambda \left( \sum_n P_n \ln (n) - C\right)$, which is maximal for $P_n=Kn^{-\lambda}$.
\subsubsection{Multiplicative Noise} An example are systems described by a multiplicative noise term in the linear Langevin equation~\cite{takayasu1997stable, biro2005power} $n_{t+1}= \gamma n_{t}+ \delta $. They can be written like $n_{t+1}=n_{t}+ h(n,t)$ where the noise term appears now as an additive term. This (e.g., Gaussian) noise term $h(n,t)$ has then $\sigma_h(n)\propto n^1$, that is, $\omega=1$. In this case, a large number of urns will attain size zero, since $p(0|n)=p_0=const$ does not decrease for larger urns, due to $\sigma(n)\propto n$. For this reason many empty urns need to be refilled and have size 1. We assume again that $s_i$ are therefore  $C=\frac{1}{M}\sum_i \ln n_i^2$ are stationary. Since the system is at constant size, we also assume $\sum_n P_n n -E$.
The Lagrangian function is $S(P) = \sum_n P_n 2 \ln P_n + \lambda \left( \sum_n P_n \ln (n) - C\right) +\beta\left(\sum_n P_n n -E\right)$, which is maximal for $P_n=Kn^{-\lambda}e^{-\beta}$. Examples in the literature often have $\alpha$ sufficiently large to not need a cutoff $\beta$ for the system to be at a given mean urn size \cite{takayasu1997stable}. An additional exit rate of urns can be added, in which case the power law exponent grows with exit rate, like in Figure \ref{fig:Simulation_E_C_beta}.

\section{Conclusions}
We have introduced a method to derive stationary distributions, by looking at them as the maximum entropy distribution of the outcomes in one iteration, for a process in discrete time. The~method provides an intuitive explanation for a size or degree distribution. It has been applied to a novel preferential attachment process for systems of constant size. The model has been analyzed for three different methods to keep the system at constant size. Each provides a realistic model for real-world applications. Results are confirmed by simulations and by summing over exact probabilities. We have also applied the method to derive the Maxwell-Boltzmann distribution for the velocity of gas particles, to the Yule process, and to multiplicative noise systems, where in each case established results are reproduced. The constraint that allowed these derivations is that the sum of entropies of the individual urns are also maximal when the system's entropy is maximal.

\vspace{6pt}

\bibliography{newbib}
\bibliographystyle{plain}

\subsection*{Author Contributions}Methodology, C.M.; formal analysis, C.M.; writing---original draft preparation, C.M.; writing---review and editing, C.C. All authors have read and agreed to the published version of the manuscript. 

\subsection*{Funding}This received funding from \url{https://www.epsrc.ac.uk/} grant numbers EP/K026003/1 and EP/L019981/1. The support of  Climate-KIC/European Institute of Innovation and Technology (ARISE project) is gratefully acknowledged.
%%%%%%%%%%%%%%%%%%%%%%%%%%%%%%%%%%%%%%%%%%
\acknowledgments{We thank Fernando Rosas, Gunnar Pruessner, Tim Evans and Chris Moore for discussion, and the two anonymous referees for their helpful comments.}
%%%%%%%%%%%%%%%%%%%%%%%%%%%%%%%%%%%%%%%%%%
\subsection*{Conflict of Interest}The authors declare no conflict of interest. 
%%\appendixtitles{yes} %Leave argument "no" if all appendix headings stay EMPTY (then no dot is printed after "Appendix A"). If the appendix sections contain a heading then change the argument to "yes".
%\appendix
\section{Supplementary material}\label{sec:supplementary}
\subsection{Size Distribution from Exact Probabilities}\label{sec:sup.discrete}
In one growth and shrink cycle, an urn of size $1$ can reach 3 possible states, 0, 1 and 2. Their~probabilities can be calculated by the probability $p_g=q$ to grow by one in the growth step, and for the following shrink step when the system has grown to $(1+q)M$, the probability to shrink by one s $p_s=\frac{1}{1+q}$. From this follows that
\begin{eqnarray}
 p(2|1)=p_g(2|1)p_s(2|2)=\label{eq:proba_nt2}\frac{q}{(1+q)^2} \bf{\equiv v} \\ %\nonumber\\
 p(1|1)=p_g(1|1)p_s(1|1)+p_g(2|1)p_s(1|2)\label{eq:proba_nt1}\\
 =\frac{1-q^2}{1+q}+\frac{2q}{(1+q)^2} =  \frac{1+q^2}{(1+q)^2}\bf{\equiv w}\nonumber\\
 p(0|1)=p_g(1|1)p_s(0|1)+p_g(2|1)p_s(0|2)\label{eq:proba_nt0}\\
 =\frac{(1-q)q}{1+q}+\frac{q^3}{(1+q)^2}= \frac{q}{(1+q)^2} \bf{\equiv v}\nonumber
\end{eqnarray}

The process can thus be rephrased as a growth process that follows $p$ given in Equations (\ref{eq:proba_nt2})--(\ref{eq:proba_nt0}), using the shorter notation $\bf v$ and $\bf w$, irrespective of the interpretation of the microfoundations. This~probability mass function has mean $m=1$ since  $2v+w=1$, and variance $Var(X)= \mathbb{E}[(X-m)^2]= v(-1)^2+w\,0^2+v\,1^2=2v$. 
For an urn of size $n$, $\mathbb{E}(X)=\mathbb{E}(X_1+X_2+\ldots+X_n)= \mathbb{E}(X_1)+\mathbb{E}(X_2)+\ldots+\mathbb{E}(X_n) =n$,  and $Var(X)=Var(X_1+X_2+\ldots+X_n)=Var(X_1)+Var (X_2)+\ldots+Var(X_n)=n2v$ and thus the standard deviation of an urn's next size $p(n_{t+1}|n_t)$ scales as
\begin{equation}
\sigma(n)\propto n^{0.5 }
\label{eq:sigma_sqrt_n}
\end{equation}
with its size $n$. 
This scaling holds whenever growth is the sum of independent growth of balls.

\begin{figure}%[H]\centering
	\begin{subfigure}[]{0.9\columnwidth}
	\includegraphics[angle=0, width=0.9\columnwidth]{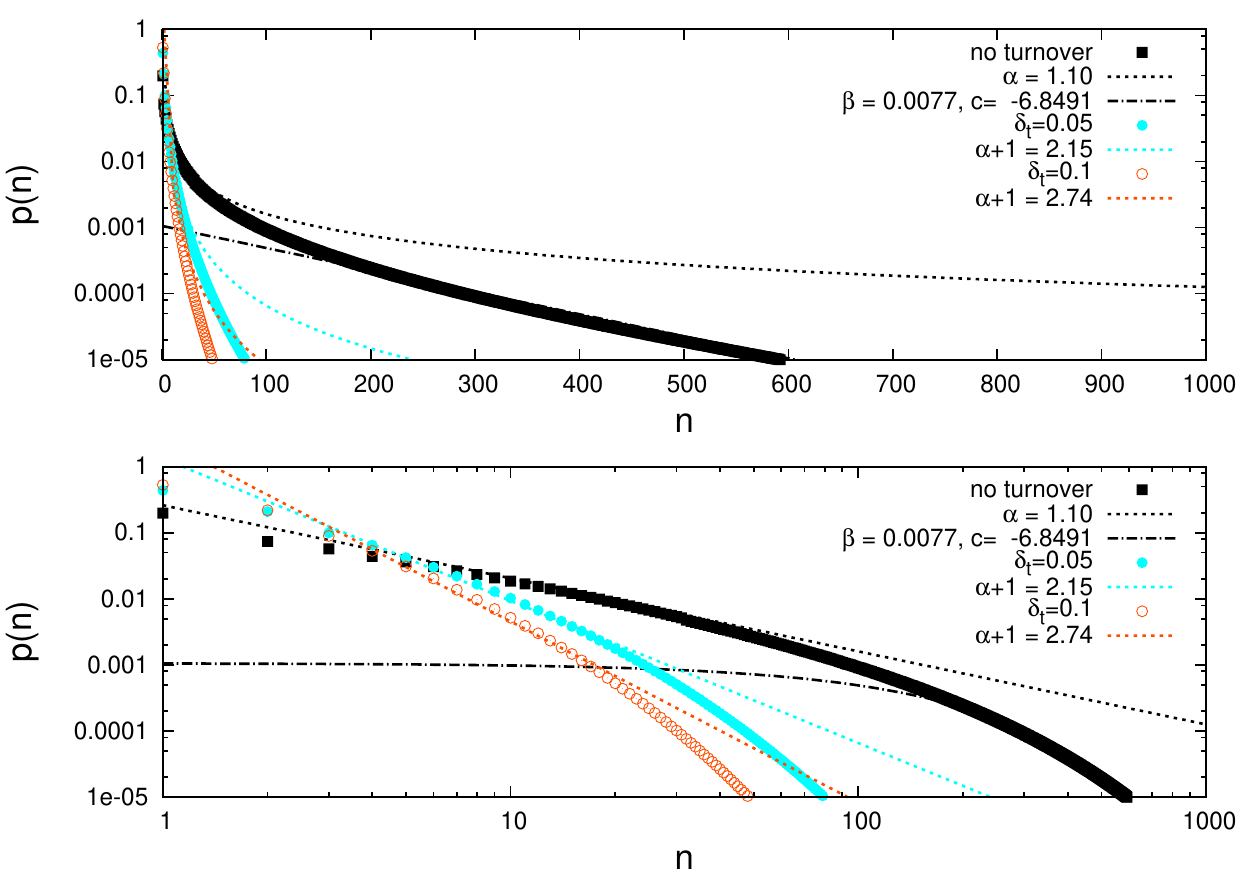}
 	\caption{}
	\end{subfigure}
	\begin{subfigure}[]{0.9\columnwidth}
	%\centering
\includegraphics[angle=0, width=0.9\columnwidth]{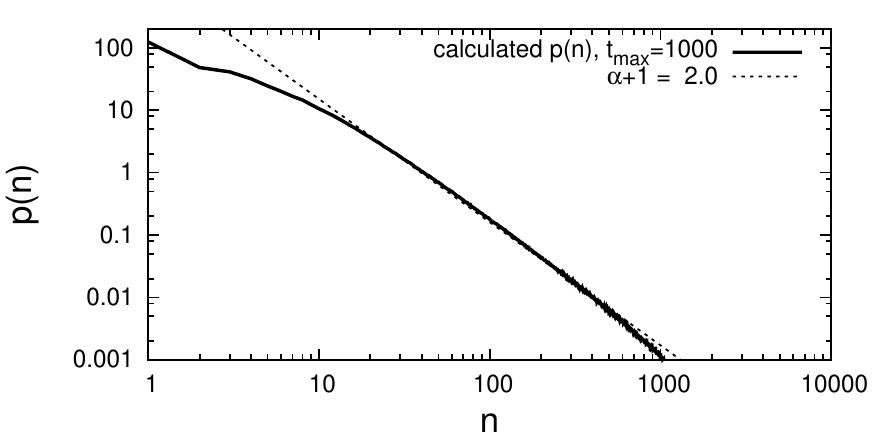}
 	\caption{}
	\label{fig:netadis_ent16gauss}
	\end{subfigure}
	\caption{ \small (\textbf{a}) Numerical normalized probability density with Gaussian $p_i$ with $\sigma(n)\propto n^{0.5}$ without and with turnover, both in log-linear and double logarithmic scale (\textbf{b}) Numerical probability density with Gaussian $p_i$ with $\sigma(n)\propto n$ generates Zipf's law $\alpha=1$.}
	\label{fig:superposition_simulation}
\end{figure}

\begin{itemize}%[leftmargin=21pt,labelsep=7pt]
\item[(i)] 
From (\ref{eq:proba_nt2})--(\ref{eq:proba_nt0}), the probabilities $p(j|k)$, can be calculated, similar to Pascal's triangle for binomial coefficients. The lowest possible $j$ for an urn of size $n_{t-1}=k$ is always $0$ (all balls leave), the largest is always $2k$ (all balls attract another ball). 
Every probability is itself a sum of terms 
\begin{equation}
p(j|k)=\sum_{(x,y)|x+y=k; y_{max}=k-|k-j|} c_{x,y,j,k}\cdot v^x(1-2v)^y
\label{eq:recursion_p}
\end{equation}

We calculated the coefficients $ \boldsymbol{c_{x,y,j,k}}$ recursively from coefficients of the corresponding addends in the 3 terms  $ p(j|k-1)$,  $ p(j-1|k-1)$ and $ p(j-2|k-1)$ with the corresponding powers $x$ and $y$:
\begin{equation}
 \boldsymbol{c_{x,y,j,k}}=\sum_{j'=j-2,j-1,j} c_{x-1, y, j', k-1}+ c_{x, y-1, j', k-1}
 \label{eq:recursion_c}
\end{equation}
if $j'$ exists, given $j' \in [0, 2(k-1)]$. The $c_{x,y,j,k}$ with $y=y_{max}$ is calculated first and no $c_{x,y,j',k-1}$ can be used in two addends for the same $(j,k)$.  
With (\ref{eq:recursion_c}) the coefficients and probabilities have been computed (until $n_{max}=1000$). Care has been taken at the implementation since (\ref{eq:recursion_p}) and (\ref{eq:recursion_c}) sum over terms of very different orders of magnitude. 
\item [(ii)]With the transition probabilities $p(j|k)$ the most probable time evolution of an urn that started at size $1$ can be calculated recursively like $p_t(n)=\sum_jp_{t-1}(j)p(n|j)$. $p_t(n=0)$ grows with $t$ and approaches $1$, since over time, the probability to have died out is increasing.
\item [(iii)]
 Assuming that equilibrium has been obtained by continuously replacing urns of size $0$ by urns of size $n=1$, the equilibrium distribution is $P_n=\frac{1}{t_{max}}\sum_t p_{n,t}$. It is shown in Figure \ref{fig:superposition_simulation}.
 \end{itemize}
 
The obtained size distribution can again be fitted by a power law with exponential cutoff (see Figure \ref{fig:superposition_simulation}).
The method applies to other processes if $p(j|k)$ can be known. We used it also for multiplicative noise systems where Zipf's law is recovered as result (Figure \ref{fig:netadis_ent16gauss}).

\vspace{1cm}

\subsection{Scheme of the Process}

\begin{figure}[H]
    \centering
    \includegraphics[width=0.9\columnwidth]{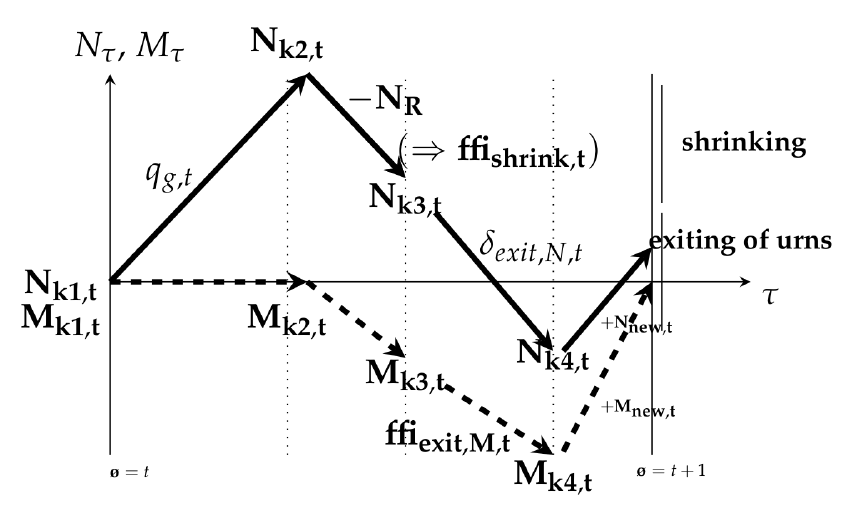}
    \caption{\small Scheme of system size evolution, in urns $M$ and balls $N$ at the different steps 1--4 within one iteration. The exit rate for balls $\delta_{exit,N,t}$ is not chosen but follows from the chosen exit rate for urns $\delta_{exit,M,t}$. $N_{k1,t+1}$ does not equal $N_{k1,t}$, since it depends on the number of balls that have exited, and on the number of balls that have been introduced when empty urns have been refilled. As  consequence, $q_{g,t}$ is adjusted such that $\left\langle N_{k2}\right\rangle$ is conserved.}
    \label{fig:scheme_iteration}
\end{figure}
\label{sec.sup.scheme}
\end{document}